# On Cascade Source Coding with A Side Information "Vending Machine"

Behzad Ahmadi and Osvaldo Simeone
CWCSPR, ECE Dept.
New Jersey Institute of Technology

Chiranjib Choudhuri and Urbashi Mitra
Ming Hsieh Dept. of Electrical Engineering
University of Southern California

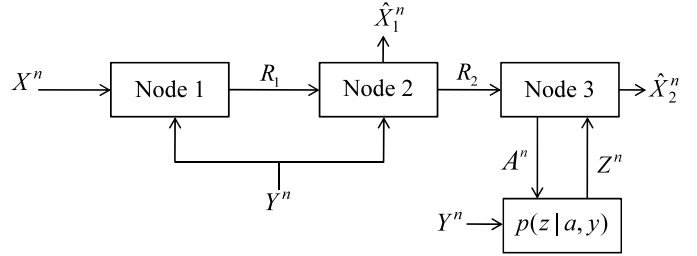

Figure 1. Cascade source coding problem with a side information "vending machine".

*Abstract*—The model of a side information "vending machine" accounts for scenarios in which acquiring side information is costly and thus should be done efficiently. In this paper, the three-node cascade source coding problem is studied under the assumption that a side information vending machine is available either at the intermediate or at the end node. In both cases, a single-letter characterization of the available trade-offs among the rate, the distortions in the reconstructions at the intermediate and at the end node, and the cost in acquiring the side information are derived under given conditions.

## I. INTRODUCTION

The concept of a side information "vending machine" has been introduced in [1] to account for source coding scenarios in which acquiring the side information at the receiver entails some cost and thus should be done efficiently. In this class of models, the quality of the side information $Y$ can be controlled by the decoder through the selection of an action $A$ that affects the effective channel between the source $X$ and the side information $Y$ via a conditional distribution $p_{Y|X,A}(y|x,a)$. Each action $A$ is associated with a cost, and the problem is that of characterizing the available trade-offs among rate, distortion and action cost.

The original paper [1] considered a point-to-point system with a single encoder and a single decoder. Various works have extended the results in [1] to multi-terminal models. Specifically, [2], [3] considered a set-up analogous to the Heegard-Berger problem [4], in which the side information may or may not be available at the decoder[1]. Instead, in [7], a specific distributed source coding setting, with two encoders and a single decoder, and a cascade source coding model, to be discussed below, were studied in the presence of a side information vending machine.

The cascade source coding model consists of three nodes arranged so that Node 1 communicates with Node 2 and Node 2 to Node 3 over finite-rate links (see Fig. 1 for an illustration of a specific example). The problem of characterizing the rate-distortion region for cascade source coding models with con-ventional side information sequences (i.e., without "vending machines") at Node 2 and Node 3 is generally open. We refer to [8] and references therein for a review of the state of the art. Two specific models for which such a characterization has been found are the settings considered in [9] and in [10], which we briefly review here for their relevance to the present work.

In [9], the cascade model in Fig. 1 was considered for the special case in which the side information sequence $Z$ at Node 3 is independent of the action sequence (so that $p(z|y,a) = p(z|y)$ in Fig. 1). In this model, the side information $Y$ measured at Node 2 is also available at Node 1, and we have the Markov chain $X - Y - Z$, so that the side information at Node 3 is degraded with respect to that of Node 2. Instead, in [10], the generalized cascade model in Fig. 2 was considered for the special case in which either rate $R_b$ or $R_1$ is zero, and the side information $Y$ at Node 2 is independent of the actions (so that $p(y|x,a) = p(y|x)$ in Fig. 2). Moreover, the reconstructions at Node 1 and Node 2 are constrained to be retrievable also by the encoder, in the sense of the Common Reconstruction (CR) requirement introduced in [6] (see below for a rigorous definition)[2].

### A. Contributions

In this paper, we extend the results in [9] and [10] to the scenarios in Fig. 1 and 2, respectively, in which the side information at Node 3, for Fig. 1, and at Node 2, for Fig. 2, are obtained through side information vending machines. In both cases, we obtain a single-letter characterization of the rate-distortion-cost trade-offs. We remark that the model of

---

[1]Reference [3] also solved the more general case in which both decoders have access to the same vending machine, and either the side information sequences produced by the vending machines at the two decoders satisfy a degradedness condition, or lossless source reconstructions are required at the decoders.

[2]A point-to-point system with action-dependent side information and the CR constraint was considered in [6].

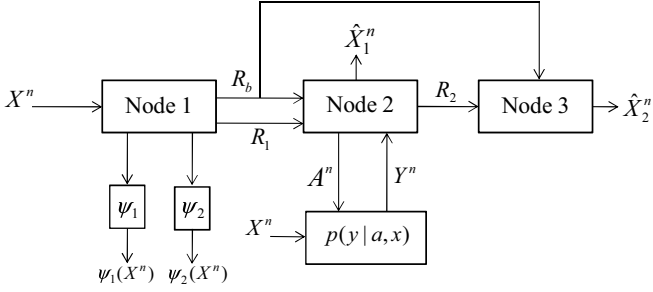

Figure 2. Cascade-broadcast source coding problem with common reconstruction constraint and a side information "vending machine".

Fig. 2 extends that of [10] also in that both rates $R_1$ and $R_b$ are allowed to be non-zero. Note, in fact, that in the model of Fig. 2, Node 1 communicates with Node 2 via a link with rate $R_1$ and broadcasts to both Node 2 and Node 3 over a link of rate $R_b$. This model is thus referred to as cascade-broadcast model in the following for short. We also remark that in [7], the characterization of the rate-distortion-cost performance for the model in Fig. 1 was obtained, but under the simplifying assumption that the side information at Node 3 be available in a causal fashion in the sense of [11].

## II. CASCADE SOURCE CODING WITH A SIDE INFORMATION VENDING MACHINE

In this section, we first describe the system model for the cascade source coding problem with a side information vending machine of Fig. 1. We then present the characterization of the corresponding rate-distortion-cost performance in Sec. II-B.

### A. System Model

The problem of cascade source coding of Fig. 1, is defined by the probability mass functions (pmfs) $p_{XY}(x,y)$ and $p_{Z|AY}(z|a,y)$ and discrete alphabets $\mathcal{X}, \mathcal{Y}, \mathcal{Z}, \mathcal{A}, \hat{\mathcal{X}}_1, \hat{\mathcal{X}}_2$, as follows. The source sequences $X^n$ and $Y^n$ with $X^n \in \mathcal{X}^n$ and $Y^n \in \mathcal{Y}^n$, respectively, are such that the pairs $(X_i, Y_i)$ for $i \in [1, n]$ are independent and identically distributed (i.i.d.) with joint pmf $p_{XY}(x, y)$. Node 1 measures sequences $X^n$ and $Y^n$ and encodes them in a message $M_1$ of $nR_1$ bits, which is delivered to Node 2. Node 2 estimates a sequence $\hat{X}_1^n \in \hat{\mathcal{X}}_1^n$ within given distortion requirements to be discussed below. Moreover, Node 2 maps the message $M_1$ received from Node 1 and the locally available sequence $Y^n$ in a message $M_2$ of $nR_2$ bits, which is delivered to node 3.

Node 3 wishes to estimate a sequence $\hat{X}_2^n \in \hat{\mathcal{X}}_2^n$ within given distortion requirements. To this end, Node 3 receives message $M_2$ and based on this, it selects an action sequence $A^n$, where $A^n \in \mathcal{A}^n$. The action sequence affects the quality of the measurement $Z^n$ of sequence $Y^n$ obtained at the Node 3. Specifically, given $A^n$ and $Y^n$, the sequence $Z^n$ is distributed as $p(z^n|a^n, y^n) = \prod_{i=1}^n p_{Z|A,Y}(z_i|y_i, a_i)$. The cost of the action sequence is defined by a cost function $\Lambda$: $\mathcal{A} \to [0, \Lambda_{\max}]$ with $0 \leq \Lambda_{\max} < \infty$, as $\Lambda(a^n) = \sum_{i=1}^n \Lambda(a_i)$.

The estimated sequence $\hat{X}_2^n$ with $\hat{X}_2^n \in \hat{\mathcal{X}}_2^n$ is then obtained as a function of $M_2$ and $Z^n$. The estimated sequences $\hat{X}_j^n$ for $j = 1, 2$ must satisfy distortion constraints defined by functions $d_j(x, \hat{x}_j)\colon \mathcal{X} \times \hat{\mathcal{X}}_j \to [0, D_{\max}]$ with $0 \leq D_{\max} < \infty$ for $j = 1, 2$, respectively. A formal description of the operations at encoder and decoder follows.

**Definition 1.** An $(n, R_1, R_2, D_1, D_2, \Gamma)$ code for the set-up of Fig. 1 consists of three source encoders, namely

$$g_1 \colon \mathcal{X}^n \times \mathcal{Y}^n \to [1, 2^{nR_1}], \quad (1)$$

which maps the sequences $X^n$ and $Y^n$ into a message $M_1$;

$$g_2 \colon \mathcal{Y}^n \times [1, 2^{nR_1}] \to [1, 2^{nR_2}], \quad (2)$$

which maps the sequence $Y^n$ and message $M_1$ into a message $M_2$; an "action" function

$$\ell \colon [1, 2^{nR_2}] \to \mathcal{A}^n, \quad (3)$$

which maps the message $M_2$ into an action sequence $A^n$; two decoders, namely

$$h_1 \colon [1, 2^{nR_1}] \times \mathcal{Y}^n \to \hat{\mathcal{X}}_1^n, \quad (4)$$

which maps the message $M_1$ and the measured sequence $Y^n$ into the estimated sequence $\hat{X}_1^n$;

$$h_2 \colon [1, 2^{nR_2}] \times \mathcal{Z}^n \to \hat{\mathcal{X}}_2^n, \quad (5)$$

which maps the message $M_2$ and the measured sequence $Z^n$ into the the estimated sequence $\hat{X}_2^n$; such that the action cost constraint $\Gamma$ and distortion constraints $D_j$ for $j = 1, 2$ are satisfied, i.e.,

$$\frac{1}{n}\sum_{i=1}^n \mathrm{E}\left[\Lambda(A_i)\right] \leq \Gamma \quad (6)$$

and $$\frac{1}{n}\sum_{i=1}^n \mathrm{E}\left[d_j(X_{ji}, h_{ji})\right] \leq D_j \text{ for } j = 1, 2, \quad (7)$$

where we have defined as $h_{1i}$ and $h_{2i}$ the $i$th symbol produced by the function $h_1(M_1, Y^n)$ and $h_2(M_2, Z^n)$, respectively.

**Definition 2.** Given a distortion-cost tuple $(D_1, D_2, \Gamma)$, a rate tuple $(R_1, R_2)$ is said to be achievable if, for any $\epsilon > 0$, and sufficiently large $n$, there exists a $(n, R_1, R_2, D_1 + \epsilon, D_2 + \epsilon, \Gamma + \epsilon)$ code.

**Definition 3.** The *rate-distortion-cost region* $\mathcal{R}(D_1, D_2, \Gamma)$ is defined as the closure of all rate tuples $(R_1, R_2)$ that are achievable given the distortion-cost tuple $(D_1, D_2, \Gamma)$.

*Remark* 4. For side information $Z$ available causally at Node 3, i.e., with decoding function (5) at Node 3 modified so that $\hat{X}_i$ is a function of $M_2$ and $Z^i$ only, the rate-distortion region $\mathcal{R}(D_1, D_2, \Gamma)$ has been derived in [7].

In the rest of this section, for simplicity of notation, we drop the subscripts from the definition of the pmfs, thus identifying a pmf by its argument.

## B. Rate-Distortion-Cost Region

In this section, a single-letter characterization of the rate-distortion region is derived.

**Proposition 5.** *The rate-distortion region $\mathcal{R}(D_1, D_2)$ for the cascade source coding problem illustrated in Fig. 1 is given by the union of all rate pairs $(R_1, R_2)$ that satisfy the conditions*

$$R_1 \geq I(X; \hat{X}_1, A, U|Y) \quad (8a)$$
$$\text{and } R_2 \geq I(X, Y; A) + I(X, Y; U|A, Z), \quad (8b)$$

*where the mutual information terms are evaluated with respect to the joint pmf*

$$p(x, y, z, a, \hat{x}_1, u) = p(x, y)p(\hat{x}_1, a, u|x, y)p(z|y, a), \quad (9)$$

*for some pmf $p(\hat{x}_1, a, u|x, y)$ such that the inequalities*

$$E[d_1(X, \hat{X}_1)] \leq D_1, \quad (10a)$$
$$E[d_2(X, f(U, Z))] \leq D_2, \quad (10b)$$
$$\text{and } E[\Lambda(A)] \leq \Gamma, \quad (10c)$$

*are satisfied for some function $f: \mathcal{U} \times \mathcal{Z} \to \hat{\mathcal{X}}_2$. Finally, $U$ is an auxiliary random variable whose alphabet cardinality can be constrained as $|\mathcal{U}| \leq |\mathcal{X}||\mathcal{Y}||\mathcal{A}| + 3$, without loss of optimality.*

*Remark* 6. For side information $Z$ independent of the action $A$ given $Y$, i.e., for $p(z|a, y) = p(z|y)$, the rate-distortion region $\mathcal{R}(D_1, D_2, \Gamma)$ in Proposition 5 reduces to that derived in [9].

The proof of the converse is provided in Appendix A. The achievability follows as a combination of the techniques proposed in [1] and [9, Theorem 1]. Here we briefly outline the main ideas, since the technical details follow from standard arguments. In the scheme at hand, Node 1 first maps sequences $X^n$ and $Y^n$ into the action sequence $A^n$ using the standard joint typicality criterion. This mapping requires a codebook of rate $I(X, Y; A)$ (see, e.g., [12, pp. 62-63]). Given the sequence $A^n$, the sequences $X^n$ and $Y^n$ are further mapped into a sequence $U^n$. This requires a codebook of size $I(X, Y; U|A)$ for each action sequence $A^n$ from standard rate-distortion considerations [12, pp. 62-63]. Similarly, given the sequences $A^n$ and $U^n$, the sequences $X^n$ and $Y^n$ are further mapped into the estimate $\hat{X}_1^n$ for Node 2 using a codebook of rate $I(X, Y; \hat{X}_1|U, A)$ for each codeword pair $(U^n, A^n)$. The thus obtained codewords are then communicated to Node 2 and Node 3 as follows.

By leveraging the side information $Y^n$ available at Node 2, conveying the codewords $A^n$, $U^n$ and $\hat{X}_1^n$ to Node 2 requires rate $I(X, Y; U, A) + I(X, Y; \hat{X}_1|U, A) - I(U, A, \hat{X}_1; Y)$ by the Wyner-Ziv theorem [12, p. 280], which equals the right-hand side of (8a). Then, sequences $A^n$ and $U^n$ are sent by Node 2 to Node 3, which requires a rate equal to the right-hand side of (8b). This follows from the rates of the used codebooks as defined above and from the Wyner-Ziv theorem, due to the side information $Z^n$ available at Node 3 upon application of the action sequence $A^n$. Finally, Node 3 produces $\hat{X}_2^n$ that leverages through a symbol-by-symbol function as $\hat{X}_{2i} = f(U_i, Z_i)$ for $i \in [1, n]$.

## III. CASCADE-BROADCAST SOURCE CODING WITH A SIDE INFORMATION VENDING MACHINE AND COMMON RECONSTRUCTION

In this section, we first describe the system model for the of cascade-broadcast source coding problem with a side information vending machine. We then present the characterization of the corresponding rate-distortion-cost performance in Sec. III-B under the CR constraint of [6]. We emphasize that, unlike the setup of Fig. 1, here the vending machine is at Node 2. Moreover, we assume that an additional broadcast link of rate $R_b$ is available that is received by Node 2 and 3. To simplify the analysis, we assume the action sequence taken by Node 2 be a function of only the broadcast message $M_b$. Finally, it is required that the reconstruction at Nodes 2 and 3 be reproducible by Node 1 as for the CR constraint [6].

### A. System Model

The problem in Fig. 2 is defined by the pmfs $p_X(x)$ and $p_{Y|AX}(y|a, x)$ and discrete alphabets $\mathcal{X}, \mathcal{Y}, \mathcal{A}, \hat{\mathcal{X}}_1, \hat{\mathcal{X}}_2$, as follows. The source sequence $X^n$ with $X^n \in \mathcal{X}^n$ is i.i.d. with pmf $p_X(x)$. Node 1 measures sequence $X^n$ and encodes it into messages $M_1$ and $M_b$ of $nR_1$ and $nR_b$ bits, respectively, which are delivered to Node 2. Moreover, message $M_b$ is broadcast also to Node 3. Node 2 estimates a sequence $\hat{X}_1^n \in \hat{\mathcal{X}}_1^n$ with distortion $D_1$. To this end, Node 2 receives messages $M_1$ and $M_b$ and based only on the latter message, it selects an action sequence $A^n$, where $A^n \in \mathcal{A}^n$. Given $A^n$ and $X^n$, the sequence $Y^n$ is distributed as $p(y^n|a^n, x^n) = \prod_{i=1}^n p_{Y|A, X}(y_i|a_i, x_i)$. The cost of the action sequence is defined as in previous section. Next, Node 2 maps messages $M_1$ and $M_b$, received from Node 1, and the locally available sequence $Y^n$ in a message $M_2$ of $nR_2$ bits, which is delivered to Node 3. Node 3 estimates a sequence $\hat{X}_2^n \in \hat{\mathcal{X}}_2^n$ as a function of $M_2$ and $M_b$ with distortion $D_2$. A formal description of the operations at encoder and decoder follows.

**Definition 7.** An $(n, R_1, R_2, R_b, D_1, D_2, \Gamma)$ code for the set-up of Fig. 2 consists of two source encoders, namely

$$g_1: \mathcal{X}^n \to [1, 2^{nR_1}] \times [1, 2^{nR_b}], \quad (11)$$

which maps the sequence $X^n$ into messages $M_1$ and $M_b$, respectively;

$$g_2: [1, 2^{nR_1}] \times [1, 2^{nR_b}] \times \mathcal{Y}^n \to [1, 2^{nR_2}] \quad (12)$$

which maps the sequence $Y^n$ and messages $(M_1, M_b)$ into a message $M_2$; an "action" function

$$\ell: [1, 2^{nR_b}] \to \mathcal{A}^n, \quad (13)$$

which maps the message $M_b$ into an action sequence $A^n$; two decoders, namely

$$h_1: [1, 2^{nR_1}] \times [1, 2^{nR_b}] \times \mathcal{Y}^n \to \hat{\mathcal{X}}_1^n, \quad (14)$$

which maps messages $M_1$ and $M_b$ and the measured sequence $Y^n$ into the estimated sequence $\hat{X}_1^n$; and

$$h_2: [1, 2^{nR_2}] \times [1, 2^{nR_b}] \to \hat{\mathcal{X}}_2, \quad (15)$$

which maps the messages $M_2$ and $M_b$ into the the estimated sequence $\hat{X}_2^n$; and two reconstruction functions

$$\psi_1: \mathcal{X}^n \to \hat{\mathcal{X}}_1^n \qquad (16a)$$
$$\text{and } \psi_2: \mathcal{X}^n \to \hat{\mathcal{X}}_2^n, \qquad (16b)$$

which map the source sequence into the estimated sequences at the encoder, namely $\psi_1(X^n)$ and $\psi_2(X^n)$, respectively; such that the action cost constraint $\Gamma$ and distortion constraints $D_j$ for $j=1,2$ are satisfied, as in (6) and (7), respectively and the CR requirements hold, namely,

$$\Pr[\psi_1(X^n) \neq h_1(M_1, M_b, Y^n)] \leq \epsilon \qquad (17a)$$
$$\Pr[\psi_2(X^n) \neq h_2(M_2, M_b)] \leq \epsilon \qquad (17b)$$

Achievable rates $(R_1, R_2, R_b)$ and rate-distortion-cost region are defined analogously to Definition 2 and Definition 3. In the rest of this section, for simplicity of notation, we drop the subscripts from the definition of the pmfs, thus identifying a pmf by its argument.

*B. Rate-Distortion-Cost Region*

In this section, a single-letter characterization of the rate-distortion region is derived.

**Proposition 8.** *The rate-distortion region $\mathcal{R}(D_1, D_2)$ for the cascade-broadcast source coding problem illustrated in Fig. 2 under the CR constraint is given by the union of all rate triples $(R_1, R_2, R_b)$ that satisfy the conditions*

$$R_b \geq I(X; A) \qquad (18a)$$
$$R_1 + R_b \geq I(X; A) + I(X; \hat{X}_1, \hat{X}_2 | A, Y) \qquad (18b)$$
$$R_2 + R_b \geq I(X; A) + I(X; \hat{X}_2 | A) \qquad (18c)$$
$$\text{and } R_1 + R_2 + R_b \geq I(X; A) + I(X; \hat{X}_2 | A) \qquad (18d)$$
$$+ I(X; \hat{X}_1 | A, Y, \hat{X}_2),$$

*where the mutual information terms are evaluated with respect to the joint pmf*

$$p(x, y, a, \hat{x}_1, \hat{x}_2) = p(x, a) p(y|x, a) p(\hat{x}_1, \hat{x}_2 | x), \qquad (19)$$

*for some pmf $p(\hat{x}_1, \hat{x}_2 | x)$ such that the inequalities*

$$E[d_j(X, \hat{X}_j)] \leq D_j, \text{ for } j = 1, 2, \qquad (20a)$$
$$\text{and } E[\Lambda(A)] \leq \Gamma, \qquad (20b)$$

*are satisfied.*

*Remark* 9. If either $R_1 = 0$ or $R_b = 0$ and the side information $Y$ is independent of the action $A$ given $X$, i.e., $p(y|a,x) = p(y|x)$, the rate-distortion region $\mathcal{R}(D_1, D_2, \Gamma)$ of Proposition 8 reduces to the one derived in [10].

The proof of converse is provided in Appendix B. The achievability follows as a combination of the techniques proposed in [1] and [10, Theorem 1]. Specifically, Node 1 first maps sequence $X^n$ into the action sequence $A^n$. This mapping requires a codebook of rate $I(X; A)$. This rate has to be conveyed over link $R_b$ by the definition of the problem and is thus received by both Node 2 and Node 3. Given the so obtained sequence $A^n$, the source sequence $X^n$ is mapped into the estimate $\hat{X}_2^n$ for Node 3 using a codebook of rate $I(X; \hat{X}_2|A)$ for each sequence $A^n$. Communicating $\hat{X}_2^n$ to Node 2 requires rate $I(X; \hat{X}_2|A, Y)$ by the Wyner-Ziv theorem. We split this rate into two rates $r_{2b}$ and $r_{2d}$, such that the message corresponding to the first rate is carried over the broadcast link of rate $R_b$ and the second on the direct link of rate $R_1$. Note that Node 2 can thus recover sequence $\hat{X}_2^n$. The rate $I(X; \hat{X}_2|A)$ of the codebook for $\hat{X}_2$ is then split into two parts, of rates $r_{0b}$ and $r_{0d}$. The message corresponding to the rate $r_{0b}$ is send to Node 3 on the broadcast link of the rate $R_b$ by Node 1, while the message of rate $r_{0d}$ is sent by Node 2 to Node 3. This way, Node 1 and Node 2 cooperate to transmit $\hat{X}_2$ to Node 3. Finally, the source sequence $X^n$ is mapped by Node 1 into the estimate $\hat{X}_1^n$ for Node 2 using a codebook of rate $I(X; \hat{X}_1|A, \hat{X}_2)$ for each pair of sequences $(A^n, \hat{X}_2^n)$. Using the Wyner-Ziv coding, this rate is reduced to $I(X; \hat{X}_1|A, Y, \hat{X}_2)$ and split into two rates $r_{1b}$ and $r_{1d}$, which are sent through links $R_b$ and $R_1$, respectively. As per discussion above, the following inequalities have to be satisfied

$$r_{0b} + r_{0d} + r_{2b} \geq I(X; \hat{X}_2 | A),$$
$$r_{2b} + r_{2d} \geq I(X; \hat{X}_2 | A, Y),$$
$$r_{1b} + r_{1d} \geq I(X; \hat{X}_1 | A, Y, \hat{X}_2),$$
$$R_1 \geq r_{1d} + r_{2d},$$
$$R_2 \geq r_{0d},$$
$$\text{and } R_b \geq r_{1b} + r_{2b} + r_{0b} + I(X; A).$$

Applying the Fourier-Motzkin algorithm [12, Appendix C] to the inequalities above, the inequalities in (18) are obtained.

## IV. CONCLUDING REMARKS

Cascade source coding models capture important aspects of multihop networks with correlated sources. This work has studied two classes of cascade models in which the quality of the available side information can be controlled via actions taken by the nodes. Many problems remain open such as characterizing the performance of the system in Fig. 1 with adaptive actions in the sense of [13] and in Fig. 2 with side information also available at Node 3.

## APPENDIX A: PROOF OF PROPOSITION 5

Here, we have the converse part of Proposition 5. For any $(n, R_1, R_2, D_1 + \epsilon, D_2 + \epsilon, \Gamma + \epsilon)$ code, we have

$$nR_1 \geq H(M_1) \geq H(M_1 | Y^n)$$
$$\stackrel{(a)}{=} I(M_1; X^n | Y^n)$$
$$\stackrel{(b)}{=} H(X^n | Y^n) - H(X^n | Y^n, M_1, M_2)$$
$$\stackrel{(c)}{=} H(X^n | Y^n) - H(X^n | Y^n, M_1, M_2, A^n)$$
$$\stackrel{(d)}{=} H(X^n | Y^n) - H(X^n | Y^n, M_1, M_2, A^n, Z^n, \hat{X}_1^n)$$
$$\stackrel{(e)}{\geq} \sum_{i=1}^{n} H(X_i | Y_i) - H(X_i | X^{i-1}, Y^i, M_2, A^i, Z^n, \hat{X}_{1i})$$
$$\stackrel{(f)}{=} \sum_{i=1}^{n} I(X_i; \hat{X}_{1i}, A_i, U_i | Y_i), \qquad (21)$$

where (*a*) follows because $M_1$ is a function of $(X^n, Y^n)$; (*b*) follows because $M_2$ is a function of $(M_1, Y^n)$; (*c*) follows because $A^n$ is a function of $M_2$; (*d*) follows since $Z^n$—$(Y^n, M_1, M_2, A^n)$—$X^n$ forms a Markov chain by the problem definition and because $\hat{X}_1^n$ is a function of $(M_1, Y^n)$; (*e*) follows since conditioning decreases entropy and since $X^n$ and $Y^n$ are i.i.d.; and (*f*) follows by defining $U_i = (M_2, X^{i-1}, Y^{i-1}, A^{i-1}, Z^{n \setminus i})$. We also have

$$nR_2 \geq H(M_2)$$
$$= I(M_2; X^n, Y^n, Z^n)$$
$$= H(X^n, Y^n, Z^n) - H(X^n, Y^n, Z^n | M_2)$$
$$= H(X^n, Y^n) + H(Z^n | X^n, Y^n) - H(Z^n | M_2)$$
$$\quad - H(X^n, Y^n | M_2, Z^n)$$
$$= \sum_{i=1}^{n} H(X_i, Y_i) + H(Z_i | Z^{i-1}, X^n, Y^n)$$
$$\quad - H(Z_i | Z^{i-1}, M_2) - H(X_i, Y_i | X^{i-1}, Y^{i-1}, M_2, Z^n)$$
$$\stackrel{(a)}{=} \sum_{i=1}^{n} H(X_i, Y_i) + H(Z_i | Z^{i-1}, X^n, Y^n, M_2, A_i)$$
$$\quad - H(Z_i | Z^{i-1}, M_2, A_i) - H(X_i, Y_i | X^{i-1}, Y^{i-1}, M_2, Z^n, A^i)$$
$$\stackrel{(b)}{\geq} \sum_{i=1}^{n} H(X_i, Y_i) + H(Z_i | X_i, Y_i, A_i) \quad (22)$$
$$\quad - H(Z_i | A_i) - H(X_i, Y_i | U_i, A_i, Z_i),$$

where (*a*) follows because $M_2$ is a function of $(M_1, Y^n)$ and thus of $(X^n, Y^n)$ and because $A^n$ is a function of $M_2$; (*b*) follows since conditioning decreases entropy, since the Markov chain relationship $Z_i$—$(X_i, Y_i, A_i)$—$(X^{n \setminus i}, Y^{n \setminus i}, M_2)$ holds and by using the definition of $U_i$.

Defining $Q$ to be a random variable uniformly distributed over $[1, n]$ and independent of all the other random variables and with $X \triangleq X_Q$, $Y \triangleq Y_Q$, $Z \triangleq Z_Q$, $A \triangleq A_Q$, $\hat{X}_1 \triangleq \hat{X}_{1Q}$, $\hat{X}_2 \triangleq \hat{X}_{2Q}$ and $U \triangleq (U_Q, Q)$, from (21) we have

$$nR_1 \geq I(X; \hat{X}_1, A, U | YQ)$$
$$\stackrel{(a)}{\geq} H(X|Y) - H(X|\hat{X}_1, A, U, Y) = I(X; \hat{X}_1, A, U|Y),$$

where in (*a*) we have used the fact that $(X^n, Y^n)$ are i.i.d and conditioning reduces entropy. Moreover, from (22) we have

$$nR_2 \geq H(X, Y|Q) + H(Z|X, Y, A, Q)$$
$$- H(Z|A, Q) - H(X, Y|U, A, Z, Q)$$
$$\stackrel{(a)}{\geq} H(X, Y) + H(Z|X, Y, A) - H(Z|A) - H(X, Y|U, A, Z)$$
$$= I(X, Y; U, A, Z) - I(Z; X, Y|A)$$
$$= I(X, Y; A) + I(X, Y; U|A, Z),$$

where (*a*) follows since $(X^n, Y^n)$ are i.i.d, since conditioning decreases entropy, by the definition of $U$ and by the problem definition. We note that the defined random variables factorizes as (9) since we have the Markov chain relationship $X$—$(A, Y)$—$Z$ by the problem definition and that $\hat{X}_2$ is a function

$f(U, Z)$ of $U$ and $Z$ by the definition of $U$. Moreover, from cost and distortion constraints (6)-(7), we have

$$D_j + \epsilon \geq \frac{1}{n} \sum_{i=1}^{n} \mathrm{E}[d_j(X_i, \hat{X}_{ji})] = \mathrm{E}[d_j(X, \hat{X}_j)], \text{ for } j = 1, 2, \tag{23a}$$

$$\text{and } \Gamma + \epsilon \geq \frac{1}{n} \sum_{i=1}^{n} \mathrm{E}[\Lambda(A_i)] = \mathrm{E}[\Lambda(A)]. \tag{23b}$$

To bound the cardinality of auxiliary random variable $U$, we fix $p(z|y, a)$ and factorize the joint pmf $p(x, y, z, a, u, \hat{x}_1)$ as

$$p(x, y, z, a, u, \hat{x}_1) = p(u)p(\hat{x}_1, a, x, y|u)p(z|y, a),$$

Therefore, for fixed $p(z|y, a)$, the quantities (8a)-(10c) can be expressed in terms of integrals $\int g_j(p(\hat{x}_1, a, x, y|u))dF(u)$, for $j = 1, ..., |\mathcal{X}||\mathcal{Y}||\mathcal{A}| + 3$, of functions $g_j(\cdot)$ that are continuous on the space of probabilities over alphabet $|\mathcal{X}| \times |\mathcal{Y}| \times |\mathcal{A}| \times |\hat{\mathcal{X}}_1|$. Specifically, we have $g_j$ for $j = 1, ..., |\mathcal{X}||\mathcal{Y}||\mathcal{A}| - 1$, given by the pmf $p(a, x, y)$ for all values of $x \in \mathcal{X}$, $y \in \mathcal{Y}$ and $a \in \mathcal{A}$, (except one), $g_{|\mathcal{X}||\mathcal{Y}||\mathcal{A}|} = H(X|A, Y, \hat{X}_1, U = u)$, $g_{|\mathcal{X}||\mathcal{Y}||\mathcal{A}|+1} = H(X, Y|A, Z, U = u)$, and $g_{|\mathcal{X}||\mathcal{Y}||\mathcal{A}|+1+j} = \mathrm{E}[d_j(X, \hat{X}_j)|U = u]$, for $j = 1, 2$. The proof in concluded by invoking Fenchel–Eggleston–Caratheodory Theorem [12, Appendix C].

APPENDIX B: PROOF OF PROPOSITION 5

Here, we prove the converse part of Proposition 5. By the CR requirements, we first observe that we have the Fano inequalities

$$H(\psi_2(X^n)|\mathrm{h}_2(M_2, M_b)) \leq n\delta(\epsilon), \tag{24a}$$
$$H(\psi_1(X^n)|\mathrm{h}_1(M_1, M_b, Y^n)) \leq n\delta(\epsilon), \tag{24b}$$

for $n$ sufficiently large, where $\delta(\epsilon)$ denotes any function such that $\delta(\epsilon) \to 0$ if $\epsilon \to 0$. Next, we have:

$$nR_b \geq H(M_b)$$
$$\stackrel{(a)}{=} I(M_b; X^n)$$
$$\stackrel{(b)}{=} H(X^n) - H(X^n|M_b, A^n)$$
$$\stackrel{(c)}{\geq} \sum_{i=1}^{n} H(X_i) - H(X_i|A_i), \tag{25}$$

where (*a*) follows because $M_b$ is a function of $X^n$; (*b*) follows because $A^n$ is a function of $M_b$; and (*c*) follows since conditioning decreases entropy and since $X^n$ is i.i.d.. In the following, for simplicity of notation, we write $\mathrm{h}_1, \mathrm{h}_2, \psi_1, \psi_2$ for the values of corresponding functions in Definition 7. Next,

we have

$$n(R_1 + R_b) \geq H(M_1, M_b)$$
$$\stackrel{(a)}{=} I(M_1, M_b; X^n, Y^n)$$
$$= H(X^n, Y^n) - H(X^n, Y^n | M_1, M_b)$$
$$= H(X^n) + H(Y^n | X^n) - H(Y^n | M_1, M_b)$$
$$- H(X^n | Y^n, M_1, M_b)$$
$$\stackrel{(b)}{=} \sum_{i=1}^n H(X_i) + H(Y_i | Y^{i-1}, X^n, M_b, A_i)$$
$$- H(Y^n | Y^{i-1}, M_1, M_b, A_i)$$
$$- H(X_i | X^{i-1}, Y^n, M_1, M_b, A_i, M_2, \mathsf{h}_1, \mathsf{h}_2)$$
$$\stackrel{(c)}{\geq} \sum_{i=1}^n H(X_i) + H(Y_i |, X_i, A_i) - H(Y_i |, A_i)$$
$$- H(X_i | Y_i, A_i, \mathsf{h}_1, \mathsf{h}_2)$$
$$= \sum_{i=1}^n I(X_i; Y_i, A_i, \mathsf{h}_1, \mathsf{h}_2) - I(Y_i; X_i | A_i)$$
$$= \sum_{i=1}^n I(X_i; Y_i, A_i, \mathsf{h}_1, \mathsf{h}_2, \psi_1, \psi_2)$$
$$- I(X_i; \psi_1, \psi_2 | Y_i, A_i, \mathsf{h}_1, \mathsf{h}_2) - I(Y_i; X_i | A_i)$$
$$\stackrel{(d)}{\geq} \sum_{i=1}^n I(X_i; Y_i, A_i, \psi_1, \psi_2) - H(\psi_1, \psi_2 | Y_i, A_i, \mathsf{h}_1, \mathsf{h}_2)$$
$$+ H(\psi_1, \psi_2 | Y_i, A_i, \mathsf{h}_1, \mathsf{h}_2, X_i) - I(Y_i; X_i | A_i)$$
$$\stackrel{(e)}{\geq} \sum_{i=1}^n I(X_i; Y_i, A_i, \psi_1, \psi_2) - I(Y_i; X_i | A_i) + n\delta(\epsilon), \quad (26)$$

where (a) follows because $(M_1, M_b)$ is a function of $X^n$; (b) follows since $X^n$ is i.i.d. and because $M_b$ is a function of $X^n$, $A_i$ is a function of $M_b$, $M_2$ is a function of $(M_1, M_b, Y^n)$ and $\mathsf{h}_1$ and $\mathsf{h}_2$ are functions of $(M_1, M_b, Y^n)$ and $(M_2, M_b)$, respectively; (c) follows since conditioning decreases entropy and because $Y_i - (X_i, A_i) - (X^{n \setminus i}, Y^{i-1}, M_b)$ forms a Markov chain; (d) follows by chain rule for mutual information and the fact that mutual information is non-negative; and (e) follows by the Fano inequality (24) and because entropy is non-negative. we can also write

$$n(R_2 + R_b) \geq H(M_2, M_b)$$
$$\stackrel{(a)}{=} I(M_2, M_b; X^n)$$
$$\stackrel{(b)}{=} H(X^n) - H(X^n | M_2, M_b, A^n, h_2)$$
$$= I(X^n; M_2, M_b, A^n, h_2, \psi_2) - I(X^n; \psi_2 | M_2, M_b, A^n, h_2)$$
$$= I(X^n; M_2, M_b, A^n, h_2, \psi_2) - H(\psi_2 | M_2, M_b, A^n, h_2)$$
$$+ H(\psi_2 | M_2, M_b, A^n, h_2)$$
$$\stackrel{(c)}{\geq} I(X^n; A^n, \psi_2) + n\delta(\epsilon)$$
$$\stackrel{(d)}{\geq} \sum_{i=1}^n H(X_i) - H(X_i | A_i, \psi_2), \quad (27)$$

where (a) follows because $(M_2, M_b)$ is a function of $X^n$; (b) follows because $A^n$ is a function of $M_a$ and $h_2$ is a function of $(M_2, M_b)$; (c) follows since entropy is non-negative and by using the Fano inequality (24a); and (d) follows because conditioning decreases entropy and since $X^n$ is i.i.d.. Moreover, with the definition $M = (M_1, M_2, M_b)$, we have the chain of inequalities

$$n(R_1 + R_2 + R_b) \geq H(M)$$
$$\stackrel{(a)}{=} H(M, A^n)$$
$$\stackrel{(b)}{=} H(A^n) + H(M | A^n)$$
$$\stackrel{(c)}{=} H(A^n) - H(A^n | X^n) + H(M | A^n) - H(M | A^n, X^n, Y^n)$$
$$\stackrel{(d)}{=} I(A^n; X^n) + I(M; X^n, Y^n | A^n)$$
$$\stackrel{(e)}{=} I(A^n; X^n) + I(M; Y^n | A^n) + I(M; X^n | Y^n, A^n)$$
$$\stackrel{(f)}{\geq} I(A^n; X^n) + I(\mathsf{h}_2; Y^n | A^n) + I(\mathsf{h}_1, \mathsf{h}_2; X^n | Y^n, A^n)$$
$$= I(A^n; X^n) + I(\mathsf{h}_2, \psi_2; Y^n | A^n) - I(\psi_2; Y^n | A^n, \mathsf{h}_2)$$
$$+ I(\mathsf{h}_1, \mathsf{h}_2, \psi_1, \psi_2; X^n | Y^n, A^n) - I(\psi_1, \psi_2; X^n | Y^n, A^n, \mathsf{h}_1, \mathsf{h}_2)$$
$$\stackrel{(g)}{\geq} I(A^n; X^n) + I(\psi_2; Z^n | A^n) - H(\psi_2 | A^n, \mathsf{h}_2)$$
$$+ H(\psi_2 | A^n, \mathsf{h}_2, Y^n) + I(\psi_1, \psi_2; X^n | Y^n, A^n)$$
$$- H(\psi_1, \psi_2 | Y^n, A^n, \mathsf{h}_1, \mathsf{h}_2) + H(\psi_1, \psi_2 | Y^n, A^n, X^n, \mathsf{h}_1, \mathsf{h}_2)$$
$$\stackrel{(h)}{\geq} I(A^n; X^n) + I(\psi_2; Y^n | A^n)$$
$$+ I(\psi_1, \psi_2; X^n | Y^n, A^n) - n\delta(\epsilon)$$
$$= I(A^n; X^n) + I(\psi_2; Y^n | A^n) + H(X^n | Y^n, A^n)$$
$$- H(X^n | Y^n, A^n, \psi_1, \psi_2) - n\delta(\epsilon), \quad (28)$$

where (a) follows because $A^n$ is a function of $M$; (b) and (e) follows by the chain rule for mutual information; (c) follows because $A^n$ is a function of $X^n$ and $M$ is a function of $(X^n, Y^n)$; (d) follows by definition of mutual information; (f) follows $\mathsf{h}_1$ and $\mathsf{h}_2$ are functions of $M$ and $(M, Y^n)$, respectively and by data processing inequality; (g) follows by chain rule for mutual information and since mutual information is non-negative; (h) follows by (24) and since entropy is non-negative. Now, we can elaborate on the first three terms in (28) as follows

$$I(A^n; X^n) + I(\psi_2; Y^n | A^n) + H(X^n | Y^n, A^n)$$
$$\stackrel{(a)}{=} H(X^n) - H(X^n | A^n) + H(X^n, Y^n | A^n)$$
$$- H(Y^n | A^n) + I(\psi_2; Y^n | A^n)$$
$$\stackrel{(b)}{=} H(X^n) - H(X^n | A^n) + H(X^n | A^n) + H(Y^n | X^n, A^n)$$
$$- H(Y^n | A^n) + I(\psi_2; Y^n | A^n)$$
$$\stackrel{(c)}{=} \sum_{i=1}^n H(X_i) + H(Y_i | Y^{i-1}, A^n, X^n) - H(Y_i | Y^{i-1}, A^n)$$
$$+ H(Y_i | Y^{i-1}, A^n) - H(Y_i | Y^{i-1}, \psi_2, A^n)$$
$$\stackrel{(d)}{\geq} \sum_{i=1}^n H(X_i) + H(Y_i | Y^{i-1}, A^n, X^n, \psi_{2i}) - H(Y_i | Y^{i-1}, \psi_{2i}, A^n)$$
$$\stackrel{(e)}{\geq} \sum_{i=1}^n H(X_i) + H(Y_i | X_i, A_i, \psi_{2i}) - H(Y_i | A_i, \psi_{2i}), \quad (29)$$

where (*a*) and (*b*) follow from the definition of mutual information and the chain rule for entropy; (*c*) follows by the chain rule for entropy and the fact that $X^n$ is i.i.d.; (*d*) follows by the fact that conditioning decreases entropy; and (*e*) follows since $Y_i$—$(A_i, X_i, \psi_{2i})$—$(X^n, Y^{i-1}, A^{n/i})$ forms a Markov chain and by defining as $\psi_{1i}$ and $\psi_{2i}$ the $i$th symbol of the sequences $\psi_1$ and $\psi_2$, respectively, and since conditioning reduces entropy. Combining (28) and (29) leads to

$$n(R_1 + R_2 + R_b) \stackrel{(a)}{\geq} \sum_{i=1}^{n} H(X_i) + H(Y_i|X_i, A_i, \psi_{2i}) \quad (30)$$
$$- H(Y_i|A_i, \psi_{2i}) - H(X_i|Y_i, A_i, \psi_{1i}, \psi_{2i}),$$

where (*a*) follows by the chain rule and by definition of mutual information.

Next, define $\hat{X}_{ji} = \psi_{ji}(X^n)$ for $j = 1, 2$ and $i = 1, 2, ..., n$ and let $Q$ be a random variable uniformly distributed over $[1, n]$ and independent of all the other random variables and with $X \triangleq X_Q$, $Y \triangleq Y_Q$, $A \triangleq A_Q$, from (25), we have

$$nR_b \geq H(X|Q) - H(X|A, Q) \stackrel{(a)}{\geq} H(X) - H(X|A) = I(X;A),$$

where (*a*) follows since $X^n$ is i.i.d. and since conditioning decreases entropy. Next, from (26), we have

$$n(R_1 + R_b) \geq I(X; Y, A, \hat{X}_1, \hat{X}_2|Q) - I(Y; X|A, Q)$$
$$\stackrel{(a)}{\geq} I(X; Y, A, \hat{X}_1, \hat{X}_2) - I(Y; X|A)$$
$$= I(X; A) + I(X; \hat{X}_1, \hat{X}_2|A, Y),$$

where (*a*) follows since $X^n$ is i.i.d., since conditioning decreases entropy and by the problem definition. From (27), we also have

$$n(R_2 + R_b) \geq H(X|Q) - H(X|A, \hat{X}_2, Q)$$
$$\stackrel{(a)}{\geq} H(X) - H(X|A, \hat{X}_2) = I(X; A, \hat{X}_2),$$

where (*a*) follows since $X^n$ is i.i.d. and by conditioning reduces entropy. Finally, from (30), we have

$$n(R_1 + R_2 + R_b) \geq H(X|Q) + H(Y|X, A, \hat{X}_2, Q)$$
$$- H(Y|A, \hat{X}_2, Q) - H(X|Y, A, \hat{X}_1, \hat{X}_2, Q)$$
$$\stackrel{(a)}{\geq} H(X) + H(Y|X, A, \hat{X}_2) - H(Y|A, \hat{X}_2) - H(X|Y, A, \hat{X}_1, \hat{X}_2)$$
$$= I(X; A, Y, \hat{X}_1, \hat{X}_2) - I(Y; X|A, \hat{X}_2)$$
$$= I(X; A) + I(X; \hat{X}_2) + I(X; \hat{X}_1|\hat{X}_2, A, Y), \quad (31)$$

where (*a*) follows since $X^n$ is i.i.d, since conditioning decreases entropy, and by the problem definition. From cost constraint (6), we have

$$\Gamma + \epsilon \geq \frac{1}{n} \sum_{i=1}^{n} \mathrm{E}\left[\Lambda(A_i)\right] = \mathrm{E}\left[\Lambda(A)\right]. \quad (32)$$

Moreover, let $\mathcal{B}$ be the event $\mathcal{B} = \{(\psi_1(X^n) \neq h_1(M_1, M_b, Y^n)) \wedge (\psi_2(X^n) \neq h_2(M_2, M_b))\}$.

Using the CR requirement (17), we have $\Pr(\mathcal{B}) \leq \epsilon$. For $j = 1, 2$, we have

$$\mathrm{E}\left[d(X_j, \hat{X}_j)\right] = \frac{1}{n} \sum_{i=1}^{n} \mathrm{E}\left[d(X_{ji}, \hat{X}_{ji})\right]$$
$$= \frac{1}{n} \sum_{i=1}^{n} \mathrm{E}\left[d(X_{ji}, \hat{X}_{ji})\Big|\mathcal{B}\right]\Pr(\mathcal{B}) + \frac{1}{n} \sum_{i=1}^{n} \mathrm{E}\left[d(X_{ji}, \hat{X}_{ji})\Big|\mathcal{B}^c\right]\Pr(\mathcal{B}^c)$$
$$\stackrel{(a)}{\leq} \frac{1}{n} \sum_{i=1}^{n} \mathrm{E}\left[d(X_{ji}, \hat{X}_{ji})\Big|\mathcal{B}^c\right]\Pr(\mathcal{B}^c) + \epsilon D_{max}$$
$$\stackrel{(b)}{\leq} \frac{1}{n} \sum_{i=1}^{n} \mathrm{E}\left[d(X_{ji}, h_{ji})\right] + \epsilon D_{max}$$
$$\stackrel{(c)}{\leq} D_j + \epsilon D_{max}, \quad (33)$$

where (*a*) follows using the fact that $\Pr(\mathcal{B}) \leq \epsilon$ and that the distortion is upper bounded by $D_{max}$; (*b*) follows by the definition of $\hat{X}_{ji}$ and $\mathcal{B}$; and (*c*) follows by (7).